# Complexity of comparison of influence of players in simple games

*Haris Aziz*

**Abstract**

Coalitional voting games appear in different forms in multi-agent systems, social choice and threshold logic. In this paper, the complexity of comparison of influence between players in coalitional voting games is characterized. The possible representations of simple games considered are simple games represented by winning coalitions, minimal winning coalitions, weighted voting game or a multiple weighted voting game. The influence of players is gauged from the viewpoint of basic player types, desirability relations and classical power indices such as Shapley-Shubik index, Banzhaf index, Holler index, Deegan-Packel index and Chow parameters. Among other results, it is shown that for a simple game represented by minimal winning coalitions, although it is easy to verify whether a player has zero or one voting power, computing the Banzhaf value of the player is #P-complete. Moreover, it is proved that multiple weighted voting games are the only representations for which it is NP-hard to verify whether the game is linear or not. For a simple game with a set $W^m$ of minimal winning coalitions and $n$ players, a $O(n.|W^m|+n^2 log(n))$ algorithm is presented which returns 'no' if the game is non-linear and returns the strict desirability ordering otherwise. The complexity of transforming simple games into compact representations is also examined.

## 1  Introduction

### 1.1  Overview

Simple games are *yes/no* coalitional voting games which arise in various mathematical contexts. Simple games were first analysed by John von Neumann and Oskar Morgenstern in their monumental book *Theory of Games and Economic Behaviour* [25]. They also examined weighted voting games in which voters have corresponding voting weights and a coalition of voters wins if their total weights equal or exceed a specified quota. Neumann and Morgenstern [25] observe that minimal winning coalitions are a useful way to represent simple games. A similar approach has been taken in [11]. We examine the complexity of computing the influence of players in simple games represented by winning coalitions, minimal winning coalitions, weighted voting games and multiple weighted voting games.

### 1.2  Outline

In Section 2, we outline different representations and properties of simple games. In Section 3, compact representations of simple games are considered. After that, the complexity of computing the influence of players in simple games is considered from the point of view of player types (Section 4), desirability ordering (Section 5), power indices and Chow parameters (Section 6). The final Section includes a summary of results and some open problems.

# 2 Background

## 2.1 Definitions

**Definitions 2.1.** A *simple voting game* is a pair $(N, v)$ with $v : 2^N \to \{0, 1\}$ where $v(\emptyset) = 0$, $v(N) = 1$ and $v(S) \leq v(T)$ whenever $S \subseteq T$. A coalition $S \subseteq N$ is *winning* if $v(S) = 1$ and *losing* if $v(S) = 0$. A simple voting game can alternatively be defined as $(N, W)$ where $W$ is the set of winning coalitions. This is called the *extensive winning form*. A minimal winning coalition (MWC) of a simple game $v$ is a winning coalition in which defection of any player makes the coalition losing. A set of minimal winning coalitions of a simple game $v$ can be denoted by $W^m(v)$. A simple voting game can be defined as $(N, W^m)$. This is called the *extensive minimal winning form*.

For the sake of brevity, we will abuse the notation to sometimes refer to game $(N, v)$ as $v$.

**Lemma 2.2.** *For a simple game $(N, W)$, $W^m$ can be computed in polynomial time.*

*Proof.* For any $S \in W$, remove elements from $S$ until any further removals would make the coalition losing. The resultant coalition $S'$ is a member of $W^m$. □

**Definition 2.3.** A coalition $S$ is *blocking* if its complement $(N \setminus S)$ is losing. For a simple game $G = (N, W)$, there is a *dual* game $G^d = (N, W^d)$ where $W^d$ contains all the blocking coalitions in $G$.

**Definitions 2.4.** The simple voting game $(N, v)$ where
$W = \{X \subseteq N, \sum_{x \in X} w_x \geq q\}$ is called a *weighted voting game*(WVG). A weighted voting game is denoted by $[q; w_1, w_2, ..., w_n]$ where $w_i$ is the voting weight of player $i$. Usually, $w_i \geq w_j$ if $i < j$.

**Definitions 2.5.** An $m$-multiple weighted voting game (MWVG) is the simple game $(N, v_1 \wedge \cdots \wedge v_m)$ where the games $(N, v_t)$ are the WVGs $[q^t; w_1^t, \ldots, w_n^t]$ for $1 \leq t \leq m$. Then $v = v_1 \wedge \cdots \wedge v_m$ is defined as:
$$v(S) = \begin{cases} 1, & \text{if } v_t(S) = 1, \forall t, 1 \leq t \leq m. \\ 0, & \text{otherwise.} \end{cases}$$
The dimension of $(N, v)$ is the least $k$ such that there exist WMGs $(N, v_1), \ldots, (N, v_k)$ such that $(N, v) = (N, v_1) \wedge \ldots \wedge (N, v_k)$.

**Definitions 2.6.** A WVG $[q; w_1, \ldots, w_n]$ is *homogeneous* if $w(S) = q$ for all $S \in W^m$. A simple game $(N, v)$ is *homogeneous* if it can be represented by a homogeneous WVG. A simple game $(N, v)$ is symmetric if $v(S) = 1$, $T \subset N$ and $|S| = |T|$ implies $v(T) = 1$.

It is easy to see that symmetric games are homogeneous with a WVG representation of $[k; \underbrace{1, \ldots, 1}_{n}]$.

That is the reason they are also called $k$-out-of-$n$ simple games.

Banzhaf index [2] and Shapley-Shubik index [23] are two classic and popular indices to gauge the voting power of players in a simple game. They are used in the context of weighted voting games, but their general definition makes them applicable to any simple game.

**Definition 2.7.** A player $i$ is *critical* in a coalition $S$ when $S \in W$ and $S \setminus i \notin W$. For each $i \in N$, we denote the number of coalitions in which $i$ is critical in game $v$ by the *Banzhaf value* $\eta_i(v)$. The *Banzhaf Index* of player $i$ in weighted voting game $v$ is $\beta_i = \frac{\eta_i(v)}{\sum_{i \in N} \eta_i(v)}$.

**Definitions 2.8.** The *Shapley-Shubik value* is the function $\kappa$ that assigns to any simple game $(N, v)$ and any voter $i$ a value $\kappa_i(v)$ where $\kappa_i = \sum_{X \subseteq N}(|X| - 1)!(n - |X|)!(v(X) - v(X - \{i\}))$. The *Shapley-Shubik index* of $i$ is the function $\phi$ defined by $\phi_i = \frac{\kappa_i}{n!}$

**Definition 2.9.** ([8]) For a simple game $v$, Chow parameters, CHOW(v) are $(|W_1|, \ldots |W_n|; |W|)$ where $W_i = \{S : S \subseteq N, i \in S\}$.

## 2.2 Desirability relation and linear games

The individual desirability relations between players in a simple game date back at least to Maschler and Peleg [18].

**Definitions 2.10.** In a simple game $(N, v)$,

- A player $i$ is *more desirable/influential* than player $j$ ($i \succeq_D j$) if $v(S \cup \{j\}) = 1 \Rightarrow v(S \cup \{i\}) = 1$ for all $S \subseteq N \setminus \{i, j\}$.

- Players $i$ and $j$ are *equally desirable/influential* or *symmetric* ($i \sim_D j$) if $v(S \cup \{j\}) = 1 \Leftrightarrow v(S \cup \{i\}) = 1$ for all $S \subseteq N \setminus \{i, j\}$.

- A player $i$ is *strictly more desirable/influential* than player $j$ ($i \succ_D j$) if $i$ is more desirable than $j$, but if $i$ and $j$ are not equally desirable.

- A player $i$ and $j$ are *incomparable* if there exist $S, T \subseteq N \setminus \{i, j\}$ such that $v(S \cup \{i\}) = 1$, $v(S \cup \{j\}) = 0$, $v(T \cup \{i\}) = 0$ and $v(T \cup \{j\}) = 1$.

Linear simple games are a natural class of simple games:

**Definitions 2.11.** A simple game is *linear* whenever the desirability relation $\succeq_D$ is complete that is any two players $i$ and $j$ are comparable ($i \succ j$, $j \succ i$ or $i \sim j$).

For linear games, the relation $R_\sim$ divides the set of voters $N$ into equivalence classes $N/R_\sim = \{N_1, \ldots, N_t\}$ such that for any $i \in N_p$ and $j \in N_q$, $i \succ j$ if and only if $p < q$.

**Definitions 2.12.** A simple game $v$ is *swap robust* if an exchange of two players from two winning coalitions cannot render both losing. A simple game is *trade robust* if any arbitrary redistributions of players in a set of winning coalitions does not result in all coalitions becoming losing.

It is easy to see that trade robustness implies swap robustness. Taylor and Zwicker [24] proved that a simple game can be represented by a WVG if and only if it is trade robust. Moreover they proved that a simple game being linear is equivalent to it being swap robust.

Taylor and Zwicker [24] show in Proposition 3.2.6 that $v$ is linear if and only if $\succ_D$ is acyclic which is equivalent to $\succ_D$ being transitive. This is not guaranteed in other desirability relations defined over coalitions [9].

**Proposition 2.13.** *A simple game with three or fewer players is linear.*

*Proof.* For a game to be non-linear, we want to player 1 and 2 to be incomparable, i.e., there exist coalitions $S_1, S_2 \subseteq N \setminus \{1, 2\}$ such that $v(\{1\} \cup S_1) = 1$, $v(\{2\} \cup S_1) = 0$, $v(\{1\} \cup S_2) = 0$ and $v(\{2\} \cup S_2) = 1$. This is clearly not possible for $n = 1$ or $2$. For $n = 3$, without loss of generality, $v$ is non-linear only if $v(\{1\} \cup \emptyset) = 1$, $v(\{2\} \cup \emptyset) = 0$, $v(\{1\} \cup \{3\}) = 0$ and $v(\{2\} \cup \{3\}) = 1$. However the fact that $v(\{1\} \cup \emptyset) = 1$ and $v(\{1\} \cup \{3\}) = 0$ leads to a contradiction. □

# 3 Compact representations

Since WVGs and MWVG are compact representations of coalitional voting games, it is natural to ask which voting games can be represented by a WVG or MWVG and what is the complexity of answering the question. Deineko and Woeginger [6] show that it is NP-hard to verify the dimension of MWVGs. We know that every WVG is linear but not every linear game has a corresponding WVG. Carreras and Freixas,[3] show that there exists a six-player simple linear game which cannot be represented by a WVG. We now define problem *X-Realizable* as the problem to decide whether game $v$ can be represented by form $X$.

**Proposition 3.1.** *WVG-Realizable is NP-hard for a MWVG.*

*Proof.* This follows directly from the proof by Deineko and Woeginger [6] that it is NP-hard to find the dimension of a MWVG. □

**Proposition 3.2.** *WVG-Realizable is in* P *for a simple game represented by its minimal winning, or winning, coalitions.*

This follows directly from Theorem 6 in [11]. The basic idea is that any simple game can be represented by linear inequalities. The idea dates back at least to [16] and the complexity of this problem was examined in the context of set covering problems. However it is one thing to know whether a simple game is WVG-Realizable and another thing to actually represent it by a WVG. It is not easy to represent a WVG-Realizable simple game by a WVG where all the weights are integers as the problem transforms from linear programming to integer programming.

**Proposition 3.3.** *(Follows from Theorem 1.7.4 of Taylor and Zwicker[11]) Any simple game is MWVG-Realizable.*

Taylor and Zwicker [24] showed that for every $n \geq 1$, there is simple game of dimension $n$. In fact it has been pointed out by Freixas and Puente [12] that that for every $n \geq 1$, there is linear simple game of dimension $n$. This shows that there is no clear relation between linearity and dimension of simple games. However it appears exceptionally hard to actually transform a simple game $(N, W)$ or $(N, W^m)$ to a corresponding MWVG. The dimension of a simple game may be exponential ($2^{(n/2)-1}$) in the number of players [24]. A simpler question is to examine the complexity of computing, or getting a bound for, the dimension of simple games.

# 4 Complexity of player types

A player in a simple game may be of various types depending on its level of influence.

**Definitions 4.1.** For a simple game $v$ on a set of players $N$, player $i$ is a

- *dummy* if and only if $\forall S \subseteq N$, if $v(S) = 1$, then $v(S \setminus \{i\}) = 1$;
- *passer* if and only if $\forall S \subseteq N$, if $i \in S$, then $v(S) = 1$;
- *vetoer* if and only if $\forall S \subseteq N$, if $i \notin S$, then $v(S) = 0$;
- *dictator* if and only if $\forall S \subseteq N$, $v(S) = 1$ if and only if $i \in S$.

It is easy to see that if a dictator exists, it is unique and all other players are dummies. This means that a dictator has voting power one, whereas all other players have zero voting power. We examine the complexity of identifying the dummy players in voting games. We already know that for the case of WVGs, Matsui and Matsui [19] proved that it is NP-hard to identify dummy players.

**Lemma 4.2.** *A player $i$ in a simple game $v$ is a dummy if and only if it is not present in any minimal winning coalition.*

*Proof.* Let us assume that player $i$ is a dummy but is present in a minimal winning coalition. That means that it is critical in the minimal winning coalition which leads to a contradiction. Now let us assume that $i$ is critical in at least one coalition $S$ such that $v(S \cup \{i\}) = 1$ and $v(S) = 0$. In that case there is a $S' \subset S$ such that $S' \cup \{i\}$ is a MWC. □

**Proposition 4.3.** *For a simple game $v$,*

1. *Dummy players can be identified in linear time if $v$ is of the form $(N, W^m)$.*

2. *Dummy players can be identified in polynomial time if $v$ is of the form $(N, W)$.*

*Proof.* We examine each case separately:

1. By Lemma 4.2, a player is a dummy if and only if it is not in member of $W^m$

2. By Lemma 2.2, $W^m$ can be computed in polynomial time.

□

From the definition, we know that a player has veto power if and only if the player is present in every winning coalition.

**Proposition 4.4.** *Vetoers can be identified in linear time for a simple game in the following representations: $(N, W)$, $(N, W^m)$, WVG and MWVG.*

*Proof.* We examine each of the cases separately:

1. $(N, W)$: Initialize all players as vetoers. For each winning coalition, if a player is not present in the coalition, remove him from the list of vetoers.

2. $(N, W^m)$: If there exists a winning coalition which does not contain player $i$, there will also exist a minimal winning coalition which does not contain $i$.

3. WVG: For each player $i$, $i$ has veto power if and only if $w(N \setminus \{i\}) < q$.

4. MWVG: For each player $i$, $i$ has veto power if and only if $N \setminus \{i\}$ is losing.

□

**Proposition 4.5.** *For a simple game represented by $(N, W)$, $(N, W^m)$, WVG or MWVG, it is easy to identify the passers and the dictator.*

*Proof.* We check both cases separately:

1. Passers: This follows from the definition of a passer. A player $i$ is a passer if and only if $v(\{i\}) = 1$.

2. Dictator: It is easy to see that if a dictator exists in a simple game, it is unique. It follows from the definition of a dictator that a player $i$ is a dictator in a simple game if $v(\{i\}) = 1$ and $v(N \setminus \{i\}) = 0$.

□

## 5 Complexity of desirability ordering

A *desirability ordering* on linear games is any ordering of players such that $1 \succeq_D 2 \succeq_D \ldots \succeq_D n$. A *strict* desirability ordering is the following ordering on players: $1 \circ 2 \circ \ldots \circ n$ where $\circ$ is either $\sim_D$ or $\succ_D$.

**Proposition 5.1.** *For a WVG:*

1. *A desirability ordering of players can be computed in polynomial time.*

2. *It is NP-hard to compute the strict desirability ordering of players.*

*Proof.* WVGs are linear games with a complete desirability ordering. For (1), it is easy to see that one desirability ordering of players in a WVG is the ordering of the weights. When $w_i = w_j$, then we know that $i \sim j$. Moreover, if $w_i > w_j$, then we know that $i$ is at least as desirable as $j$, that is $i \succeq j$. For (2), the result immediately follows from the result by Matsui and Matsui [19] where they prove that it is NP-hard to check whether two players are symmetric. □

Let $v$ be a MWVG of $m$ WVGs on $n$ players. It is easy to see that if there is an ordering of players such that such that $w_1^t \geq w_2^t \geq \ldots \geq w_n^t$ for all $t$, then $v$ is linear. However, if an ordering like this does not exist, this does not imply that the game is not linear. The following is an example of a small non-linear MWVG:

**Example 5.2.** In game $v = [10; 10, 9, 1, 0] \wedge [10; 9, 10, 0, 1]$, players 1 and 2 are incomparable. So, whereas simple games with 3 players are linear, it is easy to construct a 4 player non-linear MWVG.

**Proposition 5.3.** *It is NP-hard to verify whether a MWVG is linear or not.*

*Proof.* We prove this by a reduction from an instance of the classical NP-hard PARTITION problem.

**Name**: PARTITION
**Instance**: A set of $k$ integer weights $A = \{a_1, \ldots, a_k\}$.
**Question**: Is it possible to partition $A$, into two subsets $P_1 \subseteq A$, $P_2 \subseteq A$ so that $P_1 \cap P_2 = \emptyset$ and $P_1 \cup P_2 = A$ and $\sum_{a_i \in P_1} a_i = \sum_{a_i \in P_2} a_i$?

Given an instance of PARTITION $\{a_1, \ldots, a_k\}$, we may as well assume that $\sum_{i=1}^k a_i$ is an even integer, $2t$ say. We can transform the instance into the multiple weighted voting $v = v_1 \wedge v_2$ where

$v_1 = [q; 20a_1, \ldots, 20a_k, 10, 9, 1, 0]$ and $v_2 = [q; 20a_1, \ldots, 20a_k, 9, 10, 0, 1]$ for $q = 10 + 20t$ and $k + 4$ is the number of players.

If $A$ is a 'no' instance of PARTITION, then we see that a subset of weights $\{20a_1, \ldots, 20a_k\}$ cannot sum to $20t$. This implies that players $k+1$, $k+2$, $k+3$, and $k+4$ are not critical for any coalition. Since players $1, \ldots, k$ have the same desirability ordering in both $v_1$ and $v_2$, $v$ is linear.

Now let us assume that $A$ is a 'yes' instance of PARTITION with a partition $(P_1, P_2)$. In that case players $k+1$, $k+2$, $k+3$, and $k+4$ are critical for certain coalitions. We see that $v(\{k+1\} \cup (\{k+4\} \cup P_1)) = 1$, $v(\{k+2\} \cup (\{k+4\} \cup P_1)) = 0$, $v(\{k+1\} \cup (\{k+3\} \cup P_1)) = 0$ and $v(\{k+2\} \cup (\{k+3\} \cup P_1)) = 1$. Therefore, players $k+1$ and $k+2$ are not comparable and $v$ is not linear. □

**Proposition 5.4.** *For a simple game $v = (N, W^m)$, it can be verified in $O(n|W^m|)$ time if $v$ is linear or not.*

*Proof.* Makino [17] proved that for a positive boolean function on $n$ variables represented by the set of all minimal true vectors $minT(f)$, it can be checked in $O(n|minT(f)|)$ whether the function is *regular* (linear) or not. Makino's algorithm CHECK-FCB takes $minT(f)$ as input and outputs 'yes' if $f$ is regular and 'no' otherwise. The proof involves encoding the minimal true vectors by a *fully condensed binary tree*. Then it follows that it can be verified in $O(n(|W^m|))$ whether a simple game $v = (N, W^m)$ is linear or not. □

**Corollary 5.5.** *For a simple game $v = (N, W)$, it can be verified in polynomial time if $v$ is linear or not.*

*Proof.* We showed earlier that $(N, W)$ can be transformed into $(N, W^m)$ in polynomial time. After that we can use Makino's method [17] to verify whether the game is linear or not. □

Muroga [20] cites Winder [26] for a result concerning comparison between boolean variables and their incidence in prime implicants of a boolean function. Hilliard [14] points out that this result can be used to check the desirability relation between players in WVG-Realizable simple games. We generalize Winder's result by proving both sides of the implications and extend Hilliard's observation to that of linear simple games.

**Proposition 5.6.** *Let $v = (N, W^m)$ be a linear simple game and let $d_{k,i} = |\{S : i \in S, S \in W^m, |S| = k\}|$. Then for two players $i$ and $j$,*

1. *$i \sim_D j$ if and only if $d_{k,i} = d_{k,j}$ for $k = 1, \ldots n$.*

2. *$i \succ_D j$ if and only if for the smallest $k$ where $d_{k,i} \neq d_{k,j}$, $d_{k,i} > d_{k,j}$.*

*Proof.*   1. ($\Rightarrow$) Let us assume $i \sim_D j$. Then by definition, $v(S \cup \{j\}) = 1 \Leftrightarrow v(S \cup \{i\}) = 1$ for all $S \subseteq N \setminus \{i, j\}$. So $S \cup \{i\} \in W^m$ if and only if $S \cup \{j\} \in W^m$. Therefore, $d_{k,i} = d_{k,j}$ for $k = 1, \ldots n$.

   ($\Leftarrow$) Let us assume that $i \nsim_D j$. Since $v$ is linear, $i$ and $j$ are comparable. Without loss of generality, we assume that $i \succ_D j$. Then there exists a coalition $S \setminus \{i, j\}$ such that $v(S \cup \{i\}) = 1$ and $v(S \cup \{j\}) = 0$ and suppose $|S| = k - 1$. If $S \cup \{i\} \in W^m$, then $d_{k,i} > d_{k,j}$. If $S \cup \{i\} \notin W^m$ then there exists $S' \subset S$ such that $S' \cup \{i\} \in W^m$. Thus there exists $k' < k$ such that $d_{k',i} > d_{k',j}$.

2. ($\Rightarrow$) Let us assume that $i \succ_D j$ and let $k'$ be the smallest integer where $d_{k',i} \neq d_{k',j}$. If $d_{k',i} < d_{k',j}$, then there exists a coalition $S$ such that $S \cup \{j\} \in W^m$, $S \cup \{i\} \notin W^m$ and $|S| = k' - 1$. $S \cup \{i\} \notin W^m$ in only two cases. The first possibility is that $v(S \cup \{i\}) = 0$, but this is not true since $i \succ_D j$. The second possibility is that there exists a coalition $S' \subset S$ such that $S' \cup \{i\} \in W^m$. But that would mean that $v(S' \cup \{i\}) = 1$ and $v(S' \cup \{j\}) = 0$. This also leads to a contradiction since $k'$ is the smallest integer where $d_{k',i} \neq d_{k',j}$.

($\Leftarrow$) Let us assume that for the smallest $k$ where $d_{k,i} \neq d_{k,j}$, $d_{k,i} > d_{k,j}$. This means there exists a coalition $S$ such that $S \cup \{i\} \in W^m$, $S \cup \{j\} \notin W^m$ and $|S| = k - 1$. This means that either $v(S \cup \{j\}) = 0$ or there exists a coalition $S' \subset S$ such that $S' \cup \{i\} \in W^m$. If $v(S \cup \{j\}) = 0$, that means $i \succ_D j$. If there exists a coalition $S' \subset S$ such that $S' \cup \{j\} \in W^m$, then $d_{k',j} > d_{k',i}$ for some $k' < k$. This leads to a contradiction. $\square$

We can use this theorem and Makino's 'CHECK-FCB' algorithm [17] to make an algorithm which takes as input a simple game $(N, W^m)$ and returns NO if the game is not linear and returns the strict desirability ordering otherwise.

**Algorithm 1** Strict-desirability-ordering-of-simple-game
---
**Input:** Simple game $v = (N, W^m)$ where $N = \{1, \ldots, n\}$ and $W^m(v) = \{S_1, \ldots, S_{|W^m|}\}$.
**Output:** NO if $v$ is not linear. Otherwise output desirability equivalence classes starting from most desirable, if, $v$ is linear.
1: $X = \text{CHECK-FCB}(W^m)$
2: **if** $X = NO$ **then**
3:   **return** $NO$
4: **else**
5:   Initialize an $n \times n$ matrix $D$ where entries $d_{i,j} = 0$ for all $i$ and $j$ in $N$
6:   **for** $i = 1$ to $|W^m|$ **do**
7:     **for** each player $x$ in $S_i$ **do**
8:       $d_{|S_i|,x} \leftarrow d_{|S_i|,x} + 1$
9:     **end for**
10:  **end for**
11:  **return** classify$(N, D, 1)$
12: **end if**

**Algorithm 2** classify

**Input:** set of integers $\text{class}_{\text{index}}$, $n \times n$ matrix $D$, integer $k$.
**Output:** subclasses.

1: **if** $k = n + 1$ or $|\text{class}_{\text{index}}| = 1$ **then**
2:     **return** $\text{class}_{\text{index}}$
3: **end if**
4: $s \leftarrow |\text{class}_{\text{index}}|$
5: mergeSort($\text{class}_{\text{index}}$) in descending order such that $i > j$ if $d_{k,i} > d_{k,j}$.
6: **for** $i = 2$ to $s$ **do**
7:     subindex $\leftarrow 1$; $\text{class}_{\text{index.subindex}} \leftarrow \text{class}_{\text{index}}[1]$
8:     **if** $d_{k,\text{class}_{\text{index}}[i]} = d_{k,\text{class}_{\text{index}}[i-1]}$ **then**
9:         $\text{class}_{\text{index.subindex}} \leftarrow \text{class}_{\text{index.subindex}} \cup \text{class}_{\text{index}}[i]$
10:     **else if** $d_{k,\text{class}_{\text{index}}[i]} < d_{k,\text{class}_{\text{index}}[i-1]}$ **then**
11:         subindex $\leftarrow$ subindex $+ 1$
12:         $\text{class}_{\text{index.subindex}} \leftarrow \{\text{class}_{\text{index}}[i]\}$
13:     **end if**
14: **end for**
15: Returnset $\leftarrow \emptyset$
16: $A \leftarrow \emptyset$
17: **for** $j = 1$ to subindex **do**
18:     $A \leftarrow$ classify($\text{class}_{\text{index.j}}, D, k+1$)
19:     Returnset $\leftarrow A \cup$ Returnset
20: **end for**
21: **return** Returnset

**Proposition 5.7.** *The time complexity of Algorithm 1 is $O(n.|W^m| + n^2 log(n))$*

*Proof.* The time complexity of $CHECK - FCB$ is $O(n.|W^m|)$. The time complexity of computing matrix $D$ is $O(\text{Max}(|W^m|, n^2))$. For each iteration, sorting of sublists requires at most $O(nlog(n))$ time. There are at most $n$ loops. Therefore the total time complexity is $O(n.|W^m|) + O(\text{Max}(|W^m|, n^2)) + O(n^2 log(n)) = O(n.|W^m| + n^2 log(n))$. □

**Corollary 5.8.** *The strict desirability ordering of players in a linear simple game $v = (N, W)$ can be computed in polynomial time.*

*Proof.* The proof follows directly from the Algorithm. Moreover, we know that the set of all winning coalitions can be transformed into a set of minimal winning coalitions in polynomial time. □

## 6 Power indices and Chow parameters

Apart from the Banzhaf and Shapley-Shubik indices, there are other indices which are also used. Both the Deegan-Packel index [5] and the Holler index [15] are based on the notion of minimal winning coalitions. Minimal winning coalitions are significant with respect to coalition formation [4]. The Holler index, $H_i$ of a player $i$ in a simple game corresponds to the Banzhaf index with one difference: only swings in minimal winning coalitions contribute towards the Holler index.

**Definitions 6.1.** We define the Holler value $M_i$ as $\{S \in W^m : i \in S\}$. The Holler index which is called the *public good index* is defined by $H_i(v) = \frac{|M_i|}{\sum_{j \in N} |M_j|}$. The Deegan Packel index for player $i$ in voting game $v$ is defined by $D_i(v) = \frac{1}{|W^m|} \sum_{S \in M_i} \frac{1}{|S|}$.

Compared to the Banzhaf index and the Shapley-Shubik index, both the Holler index and the Deegan-Packel index do not always satisfy the monotonicity condition. In [19], Matsui and Matsui prove that it is NP-hard to compute the Banzhaf index, Shapley-Shubik index and Deegan-Packel index of a player. We can use a similar technique to also prove that it is NP-hard to compute the Holler index of players in a WVG. This follows directly from the fact that it is NP-hard to decide whether a player is dummy or not. Prasad and Kelly [21] and Deng and Papadimitriou [7] proved that for WVGs, computing the Banzhaf values and Shapley-Shubik values is #P-parsimonious-complete and #P-metric-complete respectively. (For details on #P-completeness and associated reductions, see [10]). Unless specified, reductions considered with #P-completeness will be Cook reductions (or polynomial-time Turing reductions).

What we see is that although it is NP-hard to compute the Holler index and Deegan-Packel of players in a WVG, the Holler index and Deegan-Packel of players in a simple game represented by its MWCs can be computed in linear time:

**Proposition 6.2.** *For a simple game $(N, W^m)$, the Holler index and Deegan-Packel index for all players can be computed in linear time.*

*Proof.* We examine each of the cases separately:

- Initialize $M_i$ to zero. Then for each $S \in W^m$, if $i \in S$, increment $M_i$ by one.
- Initialize $d_i$ to zero. Then for each $S \in W^m$, if $i \in S$, increment $d_i$, by $\frac{1}{|S|}$. Then $D_i = \frac{d_i}{|W^m|}$.

□

**Proposition 6.3.** *For a simple game $v = (N, W)$, the Banzhaf index, Shapley Shubik index, Holler index and Deegan-Packel index can be computed in polynomial time.*

*Proof.* The proof follows from the definitions. We examine each of the cases separately:

- Holler index: Transform $W$ into $W^m$ and then compute the Holler indices.
- Deegan-Packel: Transform $W$ into $W^m$ and then compute the Deegan-Packel indices.
- Banzhaf index: Initialize Banzhaf values of all players to zero. For each $S \in W$, check if the removal of a player results in $S$ becoming losing (not a member of $W$). In that case increment the Banzhaf value of that player by one.
- Shapley-Shubik index: Initialize Shapley values of all players to zero. For each $S \in W$, check if the removal of a player results in $S$ becoming losing (not a member of $W$). In that case increment the Shapley value of the player by $(|S|-1)!(n-|S|)!$.

The time complexity for all cases is polynomial in the order of the input. □

For a simple game $(N, W^m)$, listing $W$ the winning coalitions may take time exponential in the number of players. For example, let there be only one minimal winning coalition $S$ which contains players $1, \ldots, \lceil n/2 \rceil$. Then the number of winning coalitions to list is exponential in the number of players. Moreover, if $|W^m| > 1$, minimal winning coalitions can have common supersets. It is shown below that for a simple game $(N, W^m)$, even counting the total number of winning coalitions is #P-complete. Moreover, whereas it is polynomial time easy to check if a player has zero voting power (a dummy) or whether it has voting power 1 (dictator), it is #P-complete to find the actual Banzhaf or Shapley-Shubik index of the player.

**Proposition 6.4.** *For a simple game $v = (N, W^m)$, the problem of computing the Banzhaf values of players is #P-complete.*

*Proof.* The problem is clearly in #P. We prove the #P-hardness of the problem by providing a reduction from the problem of computing $|W|$. Ball and Provan [1] proved that computing $|W|$ is #P-complete. Their proof is in context of reliability functions so we first give the proof in terms of simple games. It is known that known [22] that counting the number of vertex covers is #P-complete (a vertex cover in a graph $G = (V, E)$ is a subset $C$ of $V$ such that every edge in $E$ has at least one endpoint in $C$). Now take a simple game $v = (N, W^m)$ where for any $S \in W^m$, $|S| = 2$. Game $v$ has a one-to-one correspondence with a graph $G = (V, E)$ such that $N = V$ and $\{i, j\} \in W^m$ if and only if $\{i, j\} \in E(G)$. In that case the total number of losing coalitions in $v$ is equal to the number of vertex covers of $G$. Therefore the total number of winning coalitions is equal to $2^n -$ (number of vertex covers of $G$) and computing $|W|$ is #P-complete.

Now we take a game $v = (N, W^m)$ and convert it into another game $v' = (N \cup \{n+1\}, W^m(v'))$ where for each $S \in W^m(v)$, $S \cup \{n+1\} \in W^m(v')$. In that case computing $|W(v)|$ is equivalent to computing the Banzhaf value of player $n+1$ in game $v'$. Therefore, computing Banzhaf values of players in games represented by MWCs is #P-hard. □

It follows from the proof that computing *power of collectivity to act*($\frac{|W|}{2^n}$) and Chow parameters for a simple game $(N, W^m)$ is #P-complete. Goldberg remarks in the conclusion of [13] that computing the Chow parameters of a WVG is #P-complete. It is easy to prove this. The problem of computing $|W|$ and $|W_i|$ for any player $i$ is in #P since a winning coalition can be verified in polynomial time. It is easy to reduce in polynomial time the counting version of the SUBSET-SUM problem to counting the number of winning coalitions. Moreover, for any WVG $v = [q; w_1, \ldots, w_n]$, $|W(v)|$ is equal to $|W_{n+1}(v')|$ where $v'$ is $[q; w_1, \ldots, w_n, 0]$. Therefore computing $|W_i|$ and $|W|$ for a WVG is #P-complete.

# 7 Conclusion

A summary of results has been listed in Table 1. A question mark indicates that the specified problem is still open. It is conjectured that computing Shapley values is #P-complete and it is NP-hard to compute Banzhaf indices for a simple game represented by $(N, W^m)$. It is found that although WVG, MWVG and even $(N, W^m)$ is a relatively compact representation of simple games, some of the important information encoded in these representations can apparently only be accessed by unraveling these representations. There is a need for a greater examination of transformations of simple games into compact representations.

Table 1: Summary of results

|  | $(N,W)$ | $(N,W^m)$ | WVG | MWVG |
|---|---|---|---|---|
| IDENTIFY-DUMMIES | P | linear | NP-hard | NP-hard |
| IDENTIFY-VETOERS | linear | linear | linear | linear |
| IDENTIFY-PASSERS | linear | linear | linear | linear |
| IDENTIFY-DICTATOR | linear | linear | linear | linear |
| CHOW PARAMETERS | linear | #P-complete | #P-complete | #P-complete |
| IS-LINEAR | P | P | (Always linear) | NP-hard |
| DESIRABILITY-ORDERING | P | P | P | NP-hard |
| STRICT-DESIRABILITY | P | P | NP-hard | NP-hard |
| BANZHAF-VALUES | P | #P-complete | #P-complete | #P-complete |
| BANZHAF-INDICES | P | ? | NP-hard | NP-hard |
| SHAPLEY-SHUBIK-VALUES | P | ? | #P-complete | #P-complete |
| SHAPLEY-SHUBIK-INDICES | P | ? | NP-hard | NP-hard |
| HOLLER-INDICES | P | linear | NP-hard | NP-hard |
| DEEGAN-PACKEL-INDICES | P | linear | NP-hard | NP-hard |

# References

[1] Michael O. Ball and J. Scott Provan. Disjoint products and efficient computation of reliability. *Oper. Res.*, 36(5):703–715, 1988.

[2] J F Banzhaf. Weighted voting doesn't work. *Rutgers Law Review*, 19:317–343, 1965.

[3] Francesc Carreras and Josep Freixas. Complete simple games. *Mathematical Social Sciences*, 32(2):139–155, October 1996.

[4] B. Bueno de Mesquita. Minimum winning coalition, in politics. *Neil J. Smelser and Paul B. Baltes, Editor(s)-in-Chief, International Encyclopedia of the Social & Behavioral Sciences*, pages 9889–9891, 2001.

[5] J. Deegan and E.W. Packel. A new index of power for simple n-person games. *International Journal of Game Theory*, 7(2):113123, 1978.

[6] Vladimir G. Deineko and Gerhard J. Woeginger. On the dimension of simple monotonic games. *European Journal of Operational Research*, 170(1):315–318, 2006.

[7] Xiaotie Deng and Christos H. Papadimitriou. On the complexity of cooperative solution concepts. *Math. Oper. Res.*, 19(2):257–266, 1994.

[8] Pradeep Dubey and Lloyd S. Shapley. Mathematical properties of the banzhaf power index. *Mathematics of Operations Research*, 4(2):99–131, 1979.

[9] Ezra Einy. The desirability relation of simple games. *Mathematical Social Sciences*, 10(2):155–168, October 1985.

[10] Piotr Faliszewski and Lane A. Hemaspaandra. The complexity of power-index comparison. In *AAIM*, pages 177–187, 2008.

[11] Josep Freixas, Xavier Molinero, Martin Olsen, and Maria Serna. The complexity of testing properties of simple games. *ArXiv e-prints*, 2008.

[12] Josep Freixas and Maria Albina Puente. Dimension of complete simple games with minimum. *European Journal of Operational Research*, 127(2):555–568, July 2008.

[13] Paul W. Goldberg. A bound on the precision required to estimate a boolean perceptron from its average satisfying assignment. *SIAM J. Discret. Math.*, 20(2):328–343, 2006.

[14] M. Hilliard. *Weighted voting theory and applications*. Tech. Report No. 609, chool of Operations Research and Industrial Engineering, Cornell University, 1983.

[15] M.J. Holler. Forming coalitions and measuring voting power. *Political Studies*, 30(2):262271, 1982.

[16] Sze-Tsen Hu. Threshold logic. *University of California Press, Berkeley and Los Angeles*, 1965.

[17] Kazuhisa Makino. A linear time algorithm for recognizing regular boolean functions. *J. Algorithms*, 43(2):155–176, 2002.

[18] M. Maschler and B. Peleg. A characterization, existence proof and dimension bounds for the kernel of a game. *Pacific J. Math*, 18(2):289–328., 1966.

[19] T. Matsui and Y. Matsui. A survey of algorithms for calculating power indices of weighted majority games. *Journal of the Operations Research Society of Japan*, 43(7186), 2000.

[20] S. Muroga. *Threshold logic and Its Applications*. Wiley Interscience, New York, 1971.

[21] K. Prasad and J. S. Kelly. NP-completeness of some problems concerning voting games. *Int. J. Game Theory*, 19(1):1–9, 1990.

[22] J. Scott Provan and Michael O. Ball. The complexity of counting cuts and of computing the probability that a graph is connected. *SIAM J. Comput.*, 12(4):777–788, 1983.

[23] L. S. Shapley. A value for n person games. *A. E. Roth,editor, The Shapley value*, page 3140, 1988.

[24] Alan Taylor and William Zwicker. *Simple Games: Desirability Relations, Trading, Pseudoweightings*. Princeton University Press, New Jersey, first edition, 1999.

[25] J. von Neumann and O. Morgenstern. *Theory of Games and Economic Behavior*. Princeton University Press, 1944.

[26] R.O. Winder. *Threshold logic, Ph.D. Thesis*. Mathematics Department, Princeton University, New Jersey, 1962.

**HARIS AZIZ**

*Department of Computer Science, University of Warwick, Coventry CV4 7AL, United Kingdom. haris.aziz@warwick.ac.uk.*